\documentclass[epj]{svjour}
\usepackage{bm,graphicx,amsmath}

\begin{document}

\title{Fluid transport at low Reynolds number with magnetically actuated 
       artificial cilia}

\author{Erik M. Gauger\inst{1} \and Matthew T. Downton\inst{2}
\and Holger Stark\inst{2}}
\institute{Department of Materials, University of Oxford, Parks Road, 
Oxford, OX1 3PH, UK 
\and
Institut f\"ur Theoretische Physik, Technische Universit\"at 
Berlin, Hardenbergstr. 36, D-10623 Berlin, Germany}
\authorrunning{E. Gauger \and M. Downton \and H. Stark}
\titlerunning{Fluid transport with magnetically actuated artificial cilia}
\date{\today}
\abstract{
By numerical modeling we investigate fluid transport in low-Reynolds-number 
flow achieved with a special elastic filament or artifical cilium attached 
to a planar surface. The filament 
is made of superparamagnetic particles linked together by DNA double strands. 
An external magnetic field induces dipolar interactions between the beads 
of the filament which provides a convenient way of actuating the cilium
in a well-controlled manner. The filament has recently been used to
successfully construct the first artificial micro-swimmer 
[R. Dreyfus \emph{at al.}, Nature\ \textbf{437}, 862 (2005)].
In our numerical study
we introduce a measure, which we call pumping performance, to quantify 
the fluid transport induced by the magnetically actuated cilium and 
identify an optimum stroke pattern of the filament. It consists of a slow 
transport stroke and
a fast recovery stroke. Our detailed parameter study also reveals that for 
sufficiently large magnetic fields the artificial cilium is mainly 
governed by the Mason number that compares frictional to magnetic forces. 
Initial studies on multi-cilia systems show that the pumping performance 
is very sensitive to the imposed phase lag between neighboring cilia, i.e., 
to the details of the initiated metachronal wave.
}

\maketitle


\section{Introduction} \label{intro}

Fluid transport and mixing on the microscopic level at low Reynolds
numbers is a fascinating problem that is at the center of a successful 
lab-on-chip technology \cite{Toonder08}. Nature has provided an ingenious 
solution to this challenge by using long elastic filaments, called 
flagella or cilia, that are actuated internally by molecular motors 
\cite{Brennen77,Linck01,Bray2001}. The resulting beating patterns,
in general three-dimensional, have to be non-reciprocal as Purcell
has taught us \cite{Purcell77}. Nature uses arrays of collectively 
beating cilia to transport mucus in the respiratory tract, fluid in the 
brain \cite{Ibanez2004}, or to propel microorganisms such as the paramecium.
During an early stage of a developing embryo, arrays of rotating cilia are 
responsible for establishing the left-right asymmetry in the placement of 
organs \cite{situsinversus}. Recently, experimental efforts
have been initiated to copy nature's successful concept by developing
biomimetic or artificial cilia that are actuated by external fields
\cite{Toonder08,Dreyfus05,Evans07} or to move fluid with the help of 
bacterial carpets \cite{Darnton04}.
On the other hand, there has been an increasing interest in recent times
in contributing to the theoretical understanding of how single cilia or 
flagella function (e.g. Refs. 
\cite{Wiggins,Gueron98,Camalet,Lowe,Roper,Manghi06,Kim06,Gauger06,Lauga07}) 
and of how their collective beating patterns, known as metachronal waves, 
occur (e.g. Refs. 
\cite{Gueron98,Kim06,Gueron97,Gueron99,Lagomarsino,Vilfan06,Lenz06,Guirao07,Elgeti08}).

Some years ago, Dreyfus \emph{et al.} introduced the first artificial
micro-swimmer \cite{Dreyfus05} based on an artificial flagellum. This
elastic filament consists of superparamagnetic micron-sized beads that
are linked together by pieces of double-stranded DNA 
\cite{Goubault03,Cohen05}. The flagellum is actuated by an oscillating 
magnetic field so that it drags an attached red-blood cell forward. 
Modeling using a continuum \cite{Roper} or a discretized \cite{Gauger06} 
approach described the behavior of the swimmer very well. 

Based on our theoretical work in Ref. \cite{Gauger06}, we explore in 
this article 
by numerical modeling
how the artificial flagellum or cilium can be employed for 
fluid transport. We attach the cilium to a surface and actuate it by an 
external magnetic field, the direction of which oscillates about the 
surface normal in an asymmetric fashion. 
To achieve fluid transport, the stroke pattern has to be asymmetric
\cite{Kim06}. We therefore introduce a slow transport stroke, where
the cilium remains nearly straight, and a fast recovery stroke, where the
cilium bends due to increased hydrodynamic friction.
A measure
for the amount of transported fluid during one beating cycle, called
pumping performance, helps us to discuss the system in detail and to 
identify the optimum conditions under which the cilium should be operated.
We also look at multi-cilia systems with a defined phase lag between
neighboring cilia and illustrate that such a phase lag is advantageous for
fluid transport. Of course, our metachronal waves are controlled by the
external field and do not occur through self-organized synchronization
of the beating cilia. There is now strong evidence that this intriguing
feature of biological cilia arrays is caused by hydrodynamic interactions
\cite{Gueron98,Kim06,Gueron97,Gueron99,Lagomarsino,Vilfan06,Lenz06,Guirao07,Elgeti08}). Similar synchronization phenomena mediated by hydrodynamic interactions
were discussed for sperm cells by Taylor \cite{Taylor51} and for helical 
flagella in Ref. \cite{Reichert05}.

The article is organized as follows.
Section \ref{sec.model} explains our modeling of the artificial cilium 
including the treatment of hydrodynamic interactions close to surfaces
and the magnetic actuation cycle of the filament. Section \ref{sec.pump} 
introduces the pumping performance as a measure for the transported fluid.
In Sections \ref{sec.single} and \ref{sec.multi} we then discuss the
pumping performance for a single cilium and for multi-cilia systems.
The article ends with a conclusion.

\section{Model} \label{sec.model}

\begin{figure}
\includegraphics[width=0.9\columnwidth]{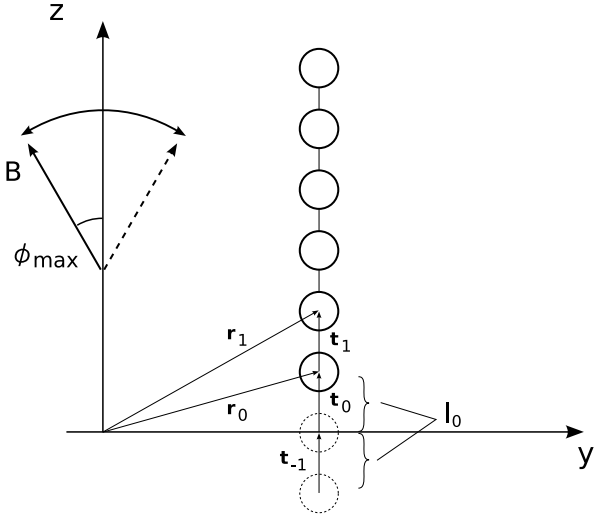}
\caption{Filament attached perpendicular to a bounding wall. The first 
three bonds $\bm{t}_{-1}$, $\bm{t}_{0}$ and $\bm{t}_{1}$ are shown. The 
actual filament starts with the bead at $\bm{r}_0$, whereas the virtual
 beads drawn as dotted lines are kept at fixed positions to anchor the 
filament perpendicular to the $xy$ plane through their elastic force 
contributions. The actuating magnetic field 
oscillating around the $z$ axis is also sketched. 
}
\label{fig1}
\end{figure}

The superparamagnetic filament is modeled by a bead-spring configuration, 
which additionally resists bending like a worm-like chain \cite{wormlike}. 
Consequently, each bead of the filament is subject to stretching and bending
forces, for which the chemical linkers are responsible, and to
dipolar interaction forces due to the induced magnetic dipoles of the beads.
We completely ignore the contributions of the chemical linkers to hydrodynamic 
friction, so the filament interacts with the fluid surrounding solely
via the hydrodynamic friction of the beads. As illustrated in Fig. \ref{fig1},
the filament is attached to a planar surface with the help of two virtual
beads that fix its position and give it 
an orientation 
orthogonal to the surface. These virtual beads contribute 
to elastic forces but do not  participate in hydrodynamic or dipolar 
interactions.

In the following, we will summarize the forces, acting on each bead 
within the filament, and the equations of motion. Details of the
derivations are given in Ref.\ \cite{Gauger06}. The equations of 
motion contain hydrodynamic mobilities which we construct up to
Rotne-Prager level based on the appropriate Green function
commonly called Blake's tensor \cite{Blake71,Happel73}. This guarantees
a vanishing fluid velocity field at the bounding surface. 
The description of the time protocol of the actuating magnetic field
concludes the section.

\subsection{Energies and forces of the superparamagnetic filament} 
\label{subsec.energies}

The total free energy for our bead-spring model is given by a sum over 
dipolar and elastic-energy contributions \cite{Gauger06},
\begin{equation}
H = H^D + H^S + H^B \, ,
\label{H_tot}
\end{equation}
where $H^D$ is the dipole-dipole interaction energy, $H^S$ is the stretching 
energy obeying Hooke's law and $H^B$ is the discretized free bending energy 
of an elastic rod. As in our previous work, the actuation and hence also the 
motion of the filament takes place in the $yz$ plane and no 
twisting of the filament occurs, allowing us to ignore a twisting energy 
in our model.  We shall briefly discuss the individual energy contributions 
of Eq.\ (\ref{H_tot}) in the following.

Let $\bm{t}_{i}$ be the vector connecting the centers of the two adjacent
beads labeled $i$ and $i-1$. A deviation $l_{i}=|\bm{t}_{i}|$ of  the beads 
from their equilibrium spacing $l_0$ then gives a total stretching free 
energy of
\begin{equation}
H^S = \frac{1}{2}  k  \sum_{i=0}^{N-1}  (l_i - l_0)^2 \, ,
\label{H_S}
\end{equation}
where $k$ is the stretching constant and $N$ the total number of beads.

Discretizing the continuum bending energy of the worm-like chain 
model \cite{wormlike,Landau91}, we obtain by a straightforward 
calculation \cite{Gauger06}
\begin{equation}
H^B = \frac{A}{l_0} \sum_{i=-1}^{N-2} (1 - \hat{\bm{t}}_{i+1} \cdot 
\hat{\bm{t}}_i) \, ,
\label{H_B}
\end{equation}
where $\hat{\bm{t}}_i = \bm{t}_i / l_{i}$ and $A$ is the bending stiffness. 

Finally, consider two identical paramagnetic beads with radius $a$ and 
magnetic susceptibility $\chi$ subject to a homogeneous external magnetic 
field $\bm{B}$. Both beads develop a dipole moment with identical orientation 
and strength,
\begin{equation}
\bm{p} = \frac{4 \pi a^3}{3 \mu_0} \chi \bm{B} \,,
\label {2.7}
\end{equation}
where $\mu_{0} = 4\pi \times 10^{-7} \mathrm{N/A^{2}}$ is the permeability of 
free space. The resulting dipoles of the beads of the filament give rise to 
the total dipole-dipole interaction energy
\begin{equation}
H^{D} = \frac{4 \pi a^{6}}{9\mu_0}(\chi B)^{2} 
\sum_ {i < j}^{N-1}\hspace*{-0.8ex} \,
 \frac{1-3(\hat{\bm{p}} \cdot \hat{\bm{r}}_{ij})}{r_{ij}^3} \, ,
\label{H_D}
\end{equation}
where $r_{ij} = |\bm{r}_{j}-\bm{r}_{i}|$, $\hat{\bm{r}}_{ij} = 
(\bm{r}_{j}-\bm{r}_{i})/r_{ij}$ and the double sum runs over all terms
with $i<j$, where the minimal $i$ is 0 and the maximal $j$ is $N-1$.

Now that all relevant interaction energies have been characterized, the 
non-hydrodynamic
force acting on each bead is obtained as usual,
\begin{equation}
\bm{F}_i  = - \bm{\nabla}_{\bm{r}_{i}} \left( H^S + H^B + H^D \right)\, ,
\label{2.2}
\end{equation}
where $\bm{\nabla}_{\bm{r}_{i}}$ is the gradient operator with respect to 
$\bm{r}_{i}$. The total 
non-hydrodynamic
force
$\bm{F}_i =  \bm{F}_j^{S} + \bm{F}_j^{B} + \bm{F}_j^{D}$ acting on bead $i$ 
is readily calculated from Eqs.\ (\ref{H_S}, \ref{H_B}, \ref{H_D}).  
Explicit expressions of $\bm{F}_i$ can also be found in Ref. \cite{Gauger06}.

\subsection{Equations of Motion}

On the micron length scale, the motion of a particle immersed in a viscous 
fluid, such as water, is entirely dominated by friction and the particle's
inertia can be neglected provided the time scale of interest 
exceeds the momentum relaxation time \cite{Dhont96}. 
Hence, the velocities $\bm{v}_{i}$ of the beads are proportional to 
the forces $\bm{F}_{j}$ acting on them and the beads obey the following 
equations of motion \cite{Dhont96}
\begin{equation}
\bm{v}_i = \sum_{j} \bm{\mu}_{ij} \bm{F}_{j} \enspace \mathrm{with} 
\enspace \bm{F}_{j} = \bm{F}_j^{S} + \bm{F}_j^{B} + \bm{F}_j^{D} \, .
\label{2.11}
\end{equation}
All the forces depend on the spatial configuration of the filament, i.~e., 
the beads' locations $\bm{r}_{i}$ . Furthermore, the dipolar forces also 
possess an explicit time dependence through the external magnetic field, 
which we use to actuate the filament.

Hydrodynamic friction enters the equations of motion via the mobilities
$\bm{\mu}_{ij}$, which depend on the geometrical configuration of the 
beads. The flow field induced by one moving bead creates a drag on all other 
beads in the vicinity and thus indirectly displaces them. Since induced flow 
fields are long ranged (they decay as $1/r$, where $r$ is the distance from 
a moving bead), hydrodynamic interactions play an important role in 
viscous systems with low Reynolds number. While hydrodynamic interactions
constitute a highly complicated many-body problem \cite{Dhont96}, their 
leading order is given by two-particle interactions. The relatively simple 
Rotne-Prager approximation can be employed whenever the beads are not so 
close together that lubrication effects need to be considered 
\cite{Dhont96,Rotne}. However, the surface, to which our filament is attached,
introduces significantly more complexity due to the no-slip boundary 
condition even when we restrict ourselves to pairwise interactions at the 
Rotne-Prager level.

\subsection{Hydrodynamic interactions}

In the low Reynolds number regime, the fluid flow velocity at an arbitrary 
point $\bm{r}$ is linearly related to a point force $\bm{F}_0$ at $\bm{r}'$ 
by \cite{Dhont96}
\begin{equation}
\bm{u}(\bm{r}) = \frac{\bm{G}(\bm{r} - \bm{r}')}{8 \pi \eta} \bm{F}_0 
\enspace ,
\label{eqn:green_equation}
\end{equation}
where $\bm{u}(\bm{r})$ denotes the flow field at $\bm{r}$ and $\eta$ is 
the fluid's viscosity. In an unbounded fluid, the Green function 
$\bm{G}(\bm{r} - \bm{r}')$, commonly referred to as Oseen tensor in 
literature, is given by 
\begin{equation}
\bm{G}(\bm{r}) = \frac{1}{r} \bm{I}+ \frac{\bm{r} \otimes \bm{r}}{r^3} 
\enspace.
\label{eqn:oseen_tensor}
\end{equation}
Here, $\bm{I}$ is the $3 \times 3$ identity matrix and $\otimes$ denotes 
the dyadic product.

The aforementioned Rotne-Prager mobilities of spherical particles are
commonly used quantities in an unbounded fluid, which can be derived
from the Oseen tensor \cite{Dhont96}. The respective self and 
cross mobilities are given by the following expressions for two spheres 
of radius $a$,
\begin{eqnarray}
\bm{\mu}_{ii} & = & \mu_{0} \bm{I} ~ , \\ 
\bm{\mu}^{rp}_{ij} & = & \mu_{0} \left\{ \frac{3}{4} \, \frac{a}{r_{ij}} 
\left[\bm{I} + \hat{\bm{r}}_{ij} \otimes \hat{\bm{r}}_{ij}
\right] \right. \nonumber \\
&  &  + \left. \frac{1}{2} \left( \frac{a}{r_{ij}} \right)^3 \left[ \bm{I} - 3 
\, \hat{\bm{r}}_{ij}  \otimes \hat{\bm{r}}_{ij} \right] \right\} \, ,
\enspace i \ne j \enspace,
\label{eqn:rotne_prager}
\end{eqnarray}
where  $\bm{r}_{ij} = \bm{r}_i - \bm{r}_j$ and $\hat{\bm{r}}_{ij} = 
\bm{r}_{ij} / r_{ij}$ and $\mu_0 = (6 \pi \eta a)^{-1}$.

However, in close proximity to a planar surface with no-slip boundary 
condition, the traditional Rotne-Prager mobilities can no longer
be employed.
The pressure and velocity fields of a point force for this boundary 
condition have been known for a long time and were first derived by 
Lorentz \cite{Lorentz1896}. Blake put these results into a modern form
replacing the Oseen tensor by the appropriate Green function \cite{Blake71}. 
The condition of a vanishing fluid velocity field 
on an infinitely extended plane is satisfied with the help of appropriate 
mirror images, similar to the image charge approach often used in 
electrostatics. In contrast to electrostatics, where it suffices to simply 
mirror the charge distribution, the hydrodynamic image system is more 
involved due to the more complicated structure of the Stokes equation 
when compared to the Poisson equation. Hence, so-called stresslet and 
source-dipole
contributions are needed in addition to the stokeslet of the
point force and its mirrored point disturbance 
(also called anti-stokeslet). This yields Blake's tensor,
\begin{eqnarray}
\bm{G}^{Blake}(\bm{r}, \bm{r}') & = & \bm{G}(\bm{r} - \bm{r}') 
+ \bm{G}^{im}(\bm{r}, \bm{\overline{r}}') \nonumber \\
& = & \bm{G}(\bm{r} - \bm{r}') \nonumber \\
& &  - \bm{G}(\bm{r}- \bm{\overline{r}}') 
+ \delta \bm{G}^{im}(\bm{r}, \bm{\overline{r}}') ~,
\label{eq:blake_tensor}
\end{eqnarray}
where $\bm{G}(\bm{r} - \bm{r}')$ and  
$\bm{G}(\bm{r} - \bm{\overline{r}}')$ are Oseen tensors,
$\bm{r}'$ is the coordinate vector of the stokeslet source, and
$\bm{\overline{r}}'$ is the position of the anti-stokeslet source, i.~e.,
the stokeslet source mirrored at the bounding $xy$ plane
[$\bm{r} = (x, y, z),  \bm{\overline{r}}' = (x', y', -z')$].
Finally, $\delta \bm{G}^{im}(\bm{r}, \bm{\overline{r}}')$ 
denotes the
source-dipole
and stresslet contributions.

For particles that are far apart from each other, it suffices to use
the Blake tensor for the mobility functions, i.~e., to treat the particles
as point-like objects, as is frequently done \cite{Vilfan06,Dufresne00}.
However, if the particles approach each other, their finite sizes become 
relevant as is the case in our filament. To take this effect into
account, we derive the mobility functions in Eq.\ (\ref{2.11}) up to the 
Rotne-Prager level following Refs. \cite{Durlofsky87,Bossis91}.
Consider $\bm{f}(\bm{r}')$ to be the force density on the surface 
$\partial V_j$ of a sphere $j$ with its centre at $\bm{r}_j$.
The force density acts on the fluid and induces the flow field
\begin{equation}
\bm{u}(\bm{r}) = \frac{1}{8 \pi \eta} \int_{\partial V_j} 
\bm{G}^{Blake}(\bm{r}, \bm{r}') \bm{f}(\bm{r}') \,dS'
\end{equation}
The leading order of $\bm{f}(\bm{r}')$ is $\bm{F}_j / (4\pi a^2)$,
where $\bm{F}_j$ denotes the total force on particle $j$.
Taking it into account and expanding $\bm{G}^{Blake}(\bm{r}, \bm{r}')$ 
about $\bm{r}' = \bm{r}_j$ yields for the flow field
\begin{equation}
\bm{u}(\bm{r}) \approx \frac{1}{8 \pi \eta} \left[ \left(1 + \frac{a^2}{6}
\bm{\nabla}_{\bm{r}'}^2 \right) \bm{G}^{Blake}(\bm{r}, \bm{r}') 
\right]_{\bm{r}' = \bm{r}_j} \bm{F}_j ~.
\label{eqn:hi-wall-flowfield}
\end{equation}
We proceed by calculating the effect of $\bm{u}(\bm{r})$ on the 
motion of another sphere $i$ with the help of Fax\'en's theorem 
\cite{Dhont96},
\begin{equation}
\bm{v}_{i}  =  \frac{1}{6 \pi \eta a} \bm{F}_i +
\left(1 + \frac{1}{6} a^2 \bm{\nabla}_{\bm{r}_i}^2 \right) 
\bm{u}(\bm{r}_i) ~. 
\label{eqn:hi-wall-faxen}
\end{equation}
Note that $\bm{u}(\bm{r}_i)$ also includes the flow fields
created by the mirror image of sphere $i$, i.e., by the term
$\bm{G}^{im}(\bm{r}, \bm{\overline{r}}')$ in the Blake 
tensor of Eq.\ (\ref{eq:blake_tensor}). Combining 
Eqs.\ (\ref{eqn:hi-wall-flowfield}) and (\ref{eqn:hi-wall-faxen}), 
we immediately obtain the cross mobilities \cite{Swan07}
\begin{equation}
\bm{\mu}_{ij} =  \frac{1}{8 \pi \eta} 
\left[ \left(1 + \frac{a^2}{6}\bm{\nabla}_{\bm{r}_i}^2 \right) 
\left(1 + \frac{a^2}{6}\bm{\nabla}_{\bm{r}_j}^2 \right) 
\bm{G}^{Blake}(\bm{r}_i, \bm{r}_j) \right] ~,
\label{eqn:crossm}
\end{equation}
and the self mobility of particle $i$,
\begin{eqnarray}
\bm{\mu}_{ii} & = &  \frac{1}{8 \pi \eta} \left[\frac{4}{3a} \bm{I} 
+ \left(1 + \frac{a^2}{6}\bm{\nabla}_{\bm{r}_i}^2 \right) \right. \nonumber \\
 & & \left.  \left(1 + \frac{a^2}{6}\bm{\nabla}_{\bm{\overline{r}}_i}^2 
\right) \bm{G}^{im}(\bm{r}_i, \bm{\overline{r}}_i) \right] ~.
\label{eqn:selfm}
\end{eqnarray}
Application of the differential operators in Eqs.\ (\ref{eqn:crossm})
and (\ref{eqn:selfm}) to the Oseen tensors $\bm{G}(\bm{r} - \bm{r}')$ 
and $\bm{G}(\bm{r}- \bm{\overline{r}}')$ in the Blake tensor 
simply gives Rotne-Prager matrices. Hence, we write the mobilities as
\begin{eqnarray}
\bm{\mu}_{ii} & = & \mu_0 \bm{I}- \bm{\mu}^{rp}(\bm{r}_i - 
\bm{\overline{r}}_i) +  \delta \bm{\mu}_{self} ~ ,
\label{selfm}
\\
\bm{\mu}_{ij} & = & \bm{\mu}^{rp}(\bm{r}_i - \bm{r}_j) - \bm{\mu}^{rp}
(\bm{r}_i - \bm{\overline{r}}_j) +  \delta \bm{\mu}_{ij} ~,
\end{eqnarray}
where $\bm{\mu}^{rp}(\bm{r})$ are the Rotne-Prager tensors defined by 
Eq. (\ref{eqn:rotne_prager}) and $\delta \bm{\mu}_{self}$ and 
$ \delta \bm{\mu}_{ij}$ 
are the sourcelet and stresslet contributions originating from 
$\delta \bm{G}^{im}$. These evaluate to more complicated expressions 
which we give in Appendix \ref{app.wallmobilities}.

\subsection{Reduced Equations of Motion}

We have shown in previous work that the number of system parameters for the 
superparamagnetic filament can be significantly reduced by introducing just 
three dimensionless variables \cite{Gauger06}. The essential parameters 
governing the dynamics of the filament were identified by rescaling the 
dynamic equations appropriately so that only reduced variables appeared. 
Instead of repeating the derivation of these reduced variables 
given in Ref.\ \cite{Gauger06}, we 
limit ourselves to discussing their physical significance.

One of the emerging quantities is
\begin{equation}
S_{p} = \left(\frac{6\pi\eta \frac{a}{l_{0}} \omega L^{4}}{A} \right)^{1/4}
= \frac{L}{l_{h}} \enspace,
\label{Sperm}
\end{equation}
which we call `sperm number' in the following \cite{Lowe}.
The physical interpretation of $S_{p}$ is that it compares bending to
frictional forces in our bead-spring chain. It bears a resemblance to 
a dimensionless quantity $S_p^{\perp}$
well-known from analytical continuum models \cite{Wiggins,Lowe} 
when substituting $6 \pi \eta a/ l_0$ by $\gamma_{\perp}$ 
which is the perpendicular friction constant per unit length of a 
slender body. Note that $6 \pi \eta a/ l_0$ approximates $\gamma_{\perp}$ 
for our bead-spring chain when hydrodynamic interactions are neglected. 
Via $S_p = L/l_{h}$, one can assign a characteristic length to the system.
In the continuum model of a slender body, the analogous quantity is
the elastohydrodynamic penetration length $l_{h}^{\perp}$ \cite{Wiggins}.
An oscillation with frequency $\omega$ started at one of the ends of a 
sufficiently long filament penetrates the filament up to the length
$l_{h}^{\perp}$. Conversely, if the filament is much shorter 
than $l_{h}^{\perp}$, i.~e., if $L \ll l_{h}^{\perp}$, the filament 
oscillates like a rigid rod over its whole length.


A second important quantity is the reduced magnetic field strength,
\begin{equation}
B_{s}  = \frac{2\pi^{1/2}a^{3}\chi N}{3\mu_{0}^{1/2}l_{0} A^{1/2}}B \, ,
\end{equation}
which more suitably describes the effect of the external field on the 
superparamagnetic filament. The number $B_{s}^{2}$ compares dipolar to 
bending forces and it is proportional to the magnetoelastic number
introduced in Ref. \cite{Dreyfus05,Roper}.

Finally, we obtain a reduced spring constant
\begin{equation}
k_{s}  = \,\, \frac{N^{2}l_{0}^{3}}{A} k \, .
\label{2.19}
\end{equation}
While $S_{p}$ and $B_{s}$ are experimentally accessible via the frequency 
$\omega$ and the magnetic field strength $B$, $k_{s}$ is a fixed quantity of 
the superparamagnetic filament itself. Therefore, we do not make explicit use 
of the stretching mode of the filament and we always choose a sufficiently 
large $k_{s}$ to keep overall length fluctuations well below 10\%. 

In what follows, we find that in the interesting regime for fluid 
transport the magnetic forces of the induced dipoles dominate over the
bending forces of the superparamagnetic filament. It therefore makes sense
to introduce a fourth dimensionless number as the ratio of frictional
to magnetic forces. In literature on magnetorheological suspensions, 
it is also called the Mason number \cite{Melle03,Cebers03}:
\begin{equation}
M_{a} = S_{p}^{4}/B_{s}^{2} \enspace.
\label{eq:Mason}
\end{equation}

\subsection{Actuation of the filament}

\begin{figure}
\includegraphics[width=0.8\columnwidth]{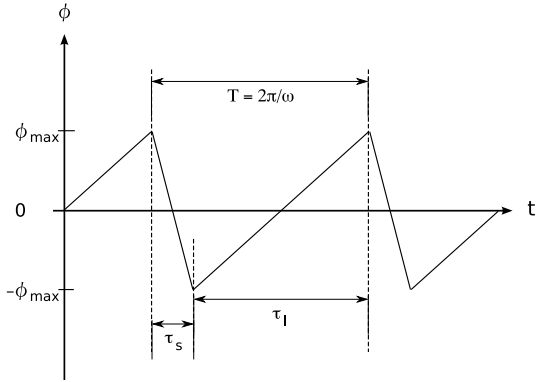}
\caption{The angle $\varphi$ enclosed by the magnetic field $\bm{B}(t)$ 
and the 
$z$ axis is shown as a function of time. $\varphi$ has different velocities 
when 
decreasing and increasing. The relative difference between 
$\tau_s$ and $\tau_l$ defines the asymmetry parameter $\varepsilon 
= (\tau_l-\tau_s)/(\tau_s+\tau_l)$.}
\label{fig:actuation}
\end{figure}

The direction of the actuating magnetic field of strength $B$ 
oscillates in the $yz$ plane (see Fig. \ref{fig:actuation}):
\begin{equation}
\bm{B}(t) = \left(0, B \sin \varphi(t), B \cos \varphi(t) \right) ~,
\end{equation}
where $\varphi(t)$ is the angle the field encloses with the  $z$ axis.
In our modeling, the angular amplitude $\varphi_{max}$ was always
$60^{\circ}$. The magnetic field induces a motion of the filament 
caused by the time-de\-pen\-dent dipole-dipole interaction of the 
filament's constituent beads.

In order to accomplish net fluid transport, a non-re\-ver\-si\-ble 
motion of the filament is required \cite{Purcell77}. In addition, 
the motion of the filament in positive and negative $y$ direction 
has to be asymmetric to achieve fluid transport along the $y$ axis.
In nature, the beating cycle of
biological cilia can typically be broken down into a power or transport
stroke, during which the cilium moves in a fashion similar to a rigid rod 
with high hydrodynamic resistance, and a recovery stroke, where it 
bends strongly and moves back with reduced resistance close to
the surface.
We mimic this 
behavior by introducing a faster stroke in one direction followed 
by a slower stroke back to the original position to complement the 
beating cycle \cite{BVariation}. However, in contrast to real cilia 
the slower stroke of our filament is the transport (`power') stroke 
and the faster stroke serves as the `recovery' with a bent filament. 
We will illustrate this below.

To accomplish the desired asymmetry, we simply rotate the $\bm{B}$ vector 
fast from one side to the other and move it back more slowly. Let 
$T = 2 \pi / \omega$ be the duration of a full actuation cycle. As 
illustrated in Fig.\ \ref{fig:actuation}, we split $T$ into two parts:
$T = \tau_s + \tau_l$. During the fast stroke with duration 
$\tau_s$, $\varphi$ is
decreasing with velocity $ - 2 \varphi_{max} / \tau_s$ and $\varphi$ is 
increasing during the slow stroke with duration $\tau_l$ with velocity 
$ 2 \varphi_{max} / \tau_l$. 
To quantify the asymmetry in the actuation cycle and therefore in the
beating pattern of the filament, we define the asymmetry parameter
\begin{equation}
\varepsilon = \frac{\tau_l-\tau_s}{\tau_s+\tau_l} \, ,
\end{equation}
which is zero for $\tau_s = \tau_l$ and tends to one in the limit of 
$\tau_l \gg \tau_s$.

In comparison with biological cilia or the work of Kim \emph{et al.} 
\cite{Kim06}, the magnetic actuation technique has the advantage of 
not requiring any active elements within the filament or its base, 
such as a driving motor with a geometrical switch for force reversal. 
Indeed, it entirely suffices to manipulate the more easily accessible 
macroscopic control field, whose time protocol allows us to achieve an 
asymmetric and non-reciprocal beating cycle.

\section{Measuring the pumping performance} \label{sec.pump}

While a non-reciprocal and asymmetric motion of the filament is required to 
generate a net fluid transport, a quantity to measure the pumping performance
of the filament is far from obvious. In the following we will introduce such
a measure based on the works in Refs. \cite{Kim06,Blake72}.

\begin{figure}
\includegraphics[width=\columnwidth]{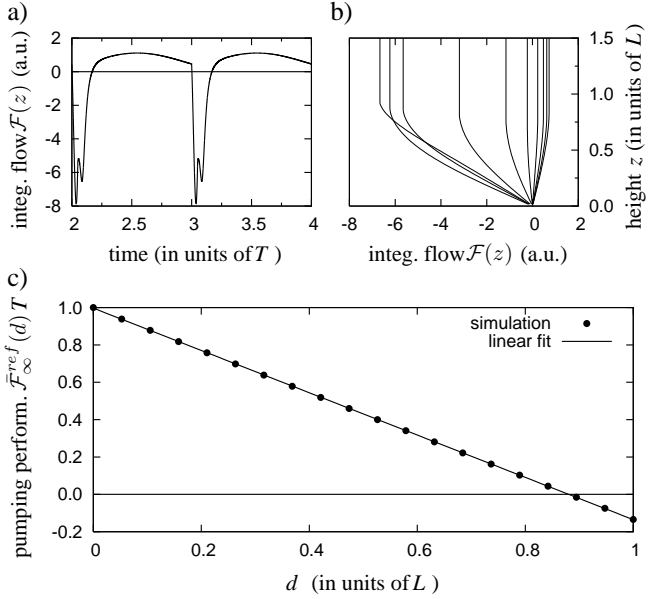}
\caption{a) Integrated flow 
$\mathcal{F}(z > L)$ 
in arbitrary units
as a function of time in units of $T$ for the parameters $S_p=3$, 
$B_s=3$ and $\varepsilon = 0.9$. 
b) Integrated flow $\mathcal{F}(z=1.5L)$ as a function of height $z$ 
at different points of time during one beating cycle. The stroboscopic 
snapshots to the right correspond to the slow part of the stroke and 
the ones to the left to the faster stroke.
c) Pumping performance $\bar{\mathcal{F}}^{ref}_{\infty}(0)$ times
period $T$ of the idealised reference cycle as a function of the 
parameter $d$ as defined by Eq. (\ref{eq:refflow}). 
The $y$ axis is scaled to units of the maximal value 
$\bar{\mathcal{F}}^{ref}_{\infty}(0) =  3.895 \times 10^{-14} ~ 
\textrm{m}^3 / \textrm{s}$ at $d = 0$.}
\label{fig:flowrefcombi}
\end{figure}

The main idea for the pumping performance is to integrate the fluid flow
initiated by the beating filament over a whole plane parallel to the
bounding surface. The integrated fluid flow is determined by the
laterally averaged Blake tensor which assumes a particularly simple 
form \cite{Blake72}:
\begin{eqnarray}
\bar{G}(z, z')	&=& \int d x d y \, G_{yy}^{Blake}(x,y,z,z') 
\nonumber \\
  &=& \frac{z+z' - \vert z -z' \vert}{2 \eta} = \frac{\min (z, z')}{\eta},
\label{eqn:laterally_avgd_gf}
\end{eqnarray}
where we restricted ourselves to the $yy$ component since fluid is
transported along the $y$ direction. Then, the integrated flow $\mathcal{F}$ 
generated by all the beads of the filament is approximated by \cite{footnote}
\begin{equation}
\mathcal{F}(z) = \frac{1}{2 \eta}  \sum_{i = 0}^{N}  
(z + z_i - \left| z - z_i\right|) F_{yi} ,
\label{eq:delF}
\end{equation}
where $z_i$ is the distance between the wall and bead $i$ and $F_{yi}$ 
is the force acting on it in $y$-direction. This expression is a function 
of the distance $z$ from the wall and
it is constant for
$z > z_i (i=1...N)$. For our beating filament, $\mathcal{F}(z)$ changes with 
time but is cyclic over a full beating cycle of the filament as is illustrated 
in Fig.\ \ref{fig:flowrefcombi}a) for $z > L$.
Figure \ref{fig:flowrefcombi}b) shows $\mathcal{F}(z)$ as a function of $z$ at 
different points of time during a typical beating cycle.  As expected,
$\mathcal{F}(z)$ only depends on $z$ roughly over the length $L$ of the 
filament and then becomes constant. Now, we consider the time average
of $\mathcal{F}(z > L)$ over one beating cycle as a suitable measure
for fluid transport, which we call pumping performance:
\begin{equation}
\bar{\mathcal{F}}_{\infty} = \frac{1}{T}\int_{t}^{t+T} d t' \, 
\mathcal{F}(z > L,t') \, ,
\label{eq:fluidflow}
\end{equation}
where $T$ is the period of one actuation cycle. 

\begin{figure}
\includegraphics[width=\columnwidth]{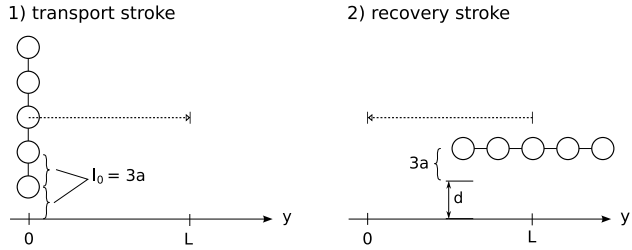}
\caption{Geometry of the idealized stroke of a rigid rod consisting
of $N$ particles of radius $a$ with distance $l_0$ between the centers 
of adjacent beads. The length parameter $L$ is defined as $L = (N-1) l_0$}
\label{fig:idealized}
\end{figure}

To understand how well the magnetically actuated filament transports fluid, 
we compare it to an idealized stroke of a rigid rod with the same
length parameter
$L = (N-1) l_0$
and the same thickness as the filament (see Fig. \ref{fig:idealized}):
\begin{enumerate}
\item In the transport stroke the rod is oriented perpendicular to the 
bounding surface and it is dragged parallel to the surface along a
distance $L$ keeping its center a distance $L/2 + 3a$ above the surface.
\item The rod is then rotated by $90^{\circ}$ to be parallel to the 
surface and then in the recovery stroke it is dragged along its long axis 
to its original position with a separation of $d+3a$ above the surface.
\end{enumerate}
In unbounded Stokes flow, the first and second part of the stroke use the
respective maximal and minimal hydrodynamic resistance to produce a net 
fluid transport over one cycle. In the presence of a no-slip wall, the 
hydrodynamic resistance increases closer to the wall. Therefore, the 
distance $d$ in the second part turns out to be an important parameter.

For the idealized stroke, we define a reference pumping performance
\begin{equation}
\bar{\mathcal{F}}^{ref}_{\infty}(d) = \frac{1}{T} \int_{t}^{t+T} d t' \, 
\mathcal{F}(z>L,t') ~ ,
\label{eq:refflow}
\end{equation}
where $T = T_{\perp} + T_{\parallel}$ is the time required for both 
parts of the cycle and the variable $d$ emphasizes the separation
between the rigid rod and the surface
for the second part of the reference cycle. We determined the reference 
pumping 
performance with the help of our filament as follows. We let a constant 
force act on each bead and numerically evaluated the average velocity of the 
beads in both parts of the strokes from which we then calculated 
the times $T_{\perp}$ and $T_{\parallel}$ the filament takes to move the
length $L$. Together with Eq. (\ref{eq:delF}) 
$\bar{\mathcal{F}}^{ref}_{\infty}(d)$ is then determined. In particular,
it becomes clear that $\bar{\mathcal{F}}^{ref}_{\infty}(d) T$ only depends
on effective friction coeffcients, $L$, and $d$. Note that in both parts 
of the reference stroke, the filament experiences an additional torque 
which alters its orientation. To really realize the idealized stroke, 
one would have to counterbalance the frictional torques appropriately.
Fig {\ref{fig:flowrefcombi}}c) shows $\bar{\mathcal{F}}^{ref}_{\infty}(d) T$
as a function of $d$. A linear fit nicely connects the data points acquired 
by simulations and represents the linear dependence of 
$\mathcal{F}(z>L,t')$ on $d$ in the recovery stroke. It makes sense that 
maximal pumping occurs when the recovery stroke of the reference cycle 
is at a minimum distance $3a$ from the wall ($d=0$) because the no-slip 
boundary condition reduces the fluid volume the rod can drag along.
Interestingly, a reversal of the pumping direction sets in at about $d=0.9$. 
Here, the perpendicular movement of the rod gives rise to less fluid 
transport compared to the parallel one simply because it is on average much 
closer to the boundary. This highlights the importance of hydrodynamic 
interactions not only amongst the beads but also with the bounding wall.

The reference stroke describes an optimum fluid transport one can
achieve with a rigid rod of length $L$ close to a surface. We
therefore compare the pumping performance $\bar{\mathcal{F}}_{\infty}$
of the magnetically actuated filament to the reference value 
$\bar{\mathcal{F}}^{ref}_{\infty}(d=0)$ at $d=0$ i.e., when the idealized 
reference cycle assumes ist maximum value. This results in the 
reduced pumping performance
\begin{equation}
\xi = \frac{\bar{\mathcal{F}}_{\infty}}{\bar{\mathcal{F}}^{ref}_{\infty}(d=0)}
~ ,
\label{eq:performance}
\end{equation}
which we use in the following. Expression $\bar{\mathcal{F}}_{\infty}$ 
in Eq. (\ref{eq:performance}) is evaluated
after the filament has assumed its limiting cycle, which typically happens 
after less than five actuation cycles after starting the simulation. 
We have checked that $\xi$ evaluates to zero
either if the motion of the filament is symmetric ($\tau_s=\tau_l$) or if it
is reciprocal, i.e., if the filament does not bend, e.g., in the case
$S_{p} \to 0$. Hence, Eq. (\ref{eq:performance}) constitutes 
a suitable measure for characterizing the pumping performance of the
magnetically actuated artificial cilium.

\section{Pumping performance of a single artificial cilium} \label{sec.single}

\begin{figure}
\includegraphics[width=\columnwidth]{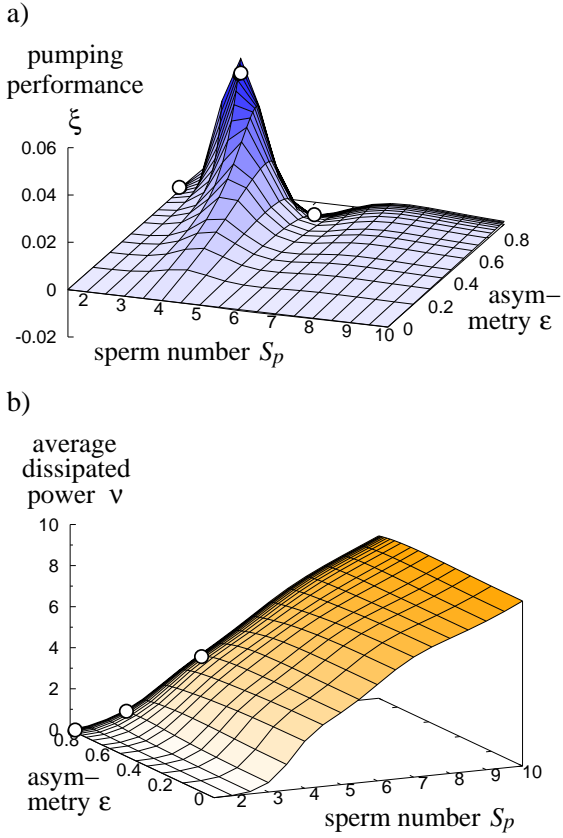}
\caption{Reduced pumping performance $\xi$ (a) and the time-averaged 
dissipated power $\nu$ in arbitrary units (b) for a single cilium as 
a function of sperm number $S_{p}$ and asymmetry parameter $\varepsilon$.
The reduced magnetic field strength is $B_s = 2.5$.
 The white dots mark parameters for which stroboscopic snapshots of the 
filament are shown in Fig. \ref{fig_snapshots}a).}
\label{fig:performance}
\end{figure}

\begin{figure}
\includegraphics[width=\columnwidth]{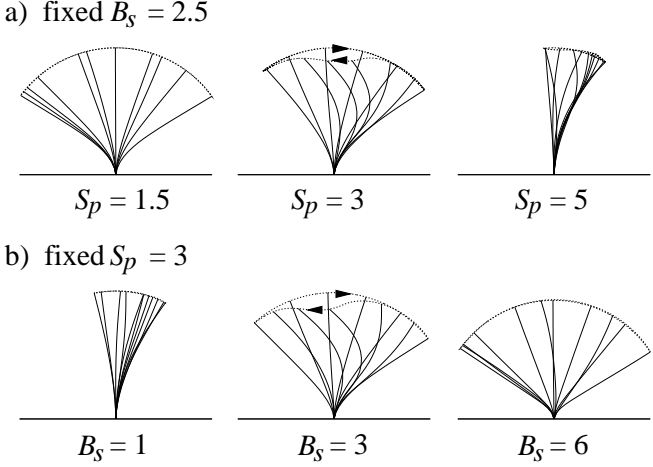}
\caption{Stroboscopic snapshots of the filament at different
times during the beating cycle for $\varepsilon = 0.9$. The trajectory of the 
top bead during one beating cycle is also indicated. In the slow transport
stroke the filament rotates clockwise, the fast recovery stroke
occurs to the left, as indicated by the arrows. A pronounced bending of the 
filament occurs only at intermediate sperm number $S_p$ and magnetic 
field strength $B_s$.}
\label{fig_snapshots}
\end{figure}

Figure \ref{fig:performance}a) shows the pumping performance $\xi$ of a 
single filament as a function of the Sperm number $S_{p}$ and the 
asymmetry parameter $\varepsilon$ at a fixed strength of the magnetic field 
$B_{s} = 2.5$. The most striking feature is that the performance is strongly 
peaked for $\varepsilon$ close to one and at $S_{p} \approx 3$. 
Such a peak is also observed in the swimming velocity of the artificial 
micro-swimmer \cite{Dreyfus05,Roper,Gauger06}.
The corresponding stroke pattern 
for $S_{p} \approx 3$
is illustrated in the middle picture of Fig. 
\ref{fig_snapshots}a). In the slow transport stroke the filament rotates
clockwise being nearly straight. It uses the high friction coefficient 
of a rigid rod dragged perpendicular to its axis to pump fluid. In the
fast recovery stroke, the filament bends due to the large hydrodynamic
friction forces that scale with velocity and then relaxes back to the
initial configuration. As Fig. \ref{fig:flowrefcombi}a) 
illustrates, fluid transport
is also noticeable in the recovery stroke. So, the
pumping performance, even for the most efficient stroke pattern, is the
result of a small asymmetry in the amount of fluid transported to the
right or left. In the example of Fig. \ref{fig:flowrefcombi}a),
which is close to the optimum stroke pattern, only 4.3 \% of the
total amount of moved fluid are effectively transported in positive
$y$ direction. As a result, the maximum pumping performance in Fig. 
\ref{fig:performance}a) is only 6\% of the reference stroke.
As expected, the pumping performance vanishes for symmetric beating
about the $z$ axis, i.e., when $\varepsilon = 0$. The same is true for 
$S_{p} \to 0$: The filament follows the actuating magnetic field 
instantaneously. Aside from the base, the filament therefore 
remains straight and the stroke is reciprocal. The left picture
in Fig. \ref{fig_snapshots}a) gives an example of such a stroke.
A reversal of the pumping direction ($\xi < 0$) occurs at $S_{p} \approx 5.5$, 
albeit only with a rather weak performance. Finally, the pumping 
performance goes to zero for increasing $S_p$ or frequency 
since the filament can no longer follow the actuating field as 
illustrated in the right picture of Fig. \ref{fig_snapshots}a). 
In summary, an optimal pumping  performances is only achieved for 
intermediate values of $S_{p}$.

Let us add a further remark. Naively, one might anticipate that the 
closer $\varepsilon$ is to one, the more pronounced the incurred bending 
during the fast part of the actuation cycle, giving rise to better 
pumping performance. However, a limit on the maximal speed with 
which the filament manages to follow the field exists. The extreme case 
of instantaneous switching of the direction of the magnetic field
in the recovery stroke ($\tau_s = 0$) is not conducive to generating fluid 
transport. On the contrary, it seems important that the free end of the 
filament follows the magnetic field's orientation during the fast stroke. 
Relaxation from the bent to a straight configuration then happens 
during the time interval $\tau_l$ allotted to the slow part of the cycle.
Consequently, the filament takes significantly longer to perform the 
actual recovery stroke of the cycle than the magnetic field. 
Naturally, this observation translates into the fact that the
$\varepsilon$ value at which maximum pumping performance occurs decreases 
with increasing $S_{p}$. In Fig. \ref{fig:performance} a), where $\varepsilon$
ranges up to 0.94 this is visible to the right of the optimal pumping 
performance.

Part b) of Fig. \ref{fig:performance} shows the time average of the 
dissipated power,
\begin{equation}
\nu = \frac{1}{T} \int_t^{t+T} \sum_{i=0}^{N} \bm{v}_{i} \cdot \bm{F}_i 
d t' ~ ,
\end{equation}
as a function of $S_p$ and $\varepsilon$. Surprisingly, it does not 
display such a pronounced behavior as the pumping performance.
Particularly, there is hardly any dependence on the asymmetry parameter
$\varepsilon$ visible. Being proportional to the square of the beads' 
velocities, one could assume $\nu \sim \omega^2$ and therefore 
$\nu \sim S_{p}^8$ (recall that  $S_{p} \sim \omega^{1/4}$). 
At small $S_p$, the data shown in Fig. \ref{fig:performance}b)
do show a steep incline. However, they do not increase as $S_{p}^8$, 
because already at $S_p$ around $2$ the filament does not completely follow
the actuating field as the snaphots in Fig. \ref{fig_snapshots}a) 
illustrate.
This effect becomes more pronounced for increasing $S_p$, and the data 
clearly deviate from the naive assumption of $\nu \sim S_{p}^8$.
The third picture in Fig. \ref{fig_snapshots}a) demonstrates that
already at $S_p=5$ the filament lags significantly behind the actuating 
magnetic field.

\begin{figure}
\includegraphics[width=\columnwidth]{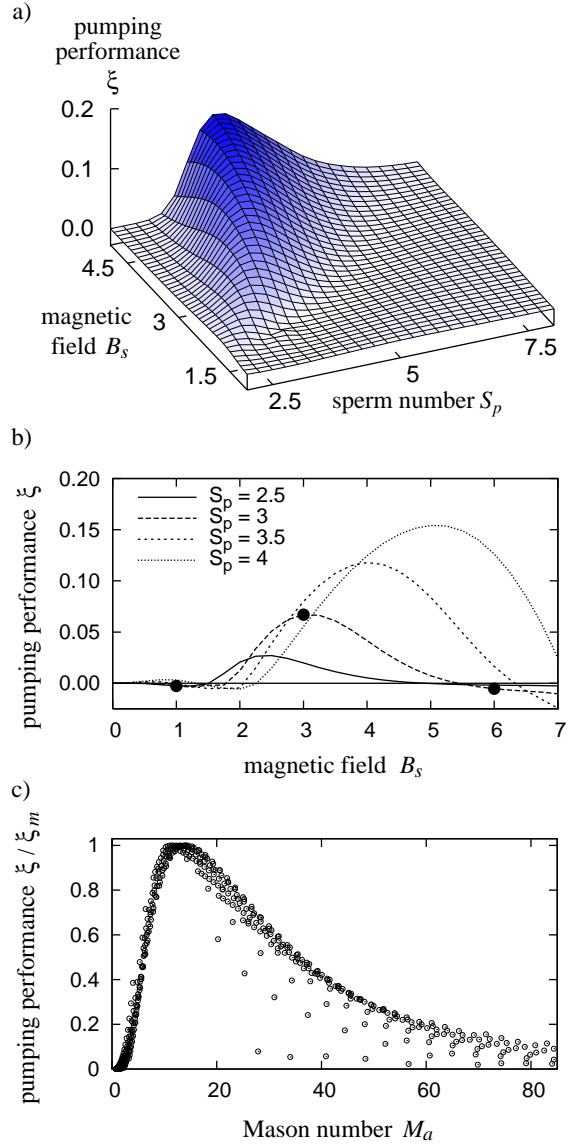}
\caption{a) Pumping performance $\xi$ as a function of $S_p$ and $B_s$ 
for $\varepsilon = 0.9$. 
b) Plot of $\xi$ versus $B_s$ for different $S_p$. 
The black dots mark parameters for which stroboscopic 
snapshots of the filament are shown in Fig. \ref{fig_snapshots}b).
c) Pumping performance $\xi$ as a function of the Mason number 
$M_a = S_p^4/B_s^2$, where $\xi$ is given in units of the maximum
values $\xi_m$ when $B_s$ is kept constant.}
\label{fig:magdepend}
\end{figure}

As illustrated by the three-dimensional plot of Fig. \ref{fig:magdepend}a),
there exists a pronounced dependence of the pumping performance $\xi$ 
on the strength $B_s$ of the actuating magnetic field which has also been
observed in the swimming velocity of the artificial micro-swimmer 
\cite{Dreyfus05,Roper,Gauger06}.
Figure \ref{fig:magdepend}b) shows this behavior for different constant values
of $S_p$. Increasing the magnetic field from zero, the pumping perfomance 
stays close to zero and then, beyond a certain threshold value, it grows 
until it reaches a maximum. It finally decreases and even becomes negative.
The snapshots in Fig. \ref{fig_snapshots}b) again explain this behavior. Small 
field strengths $B_s$ are not high enough to overcome the hydrodynamic 
friction forces and therefore the motion of the filament is very 
limited. On the other hand, at large strengths $B_s$ the filament
is always straight and therefore performs a reciprocal motion.
An optimal stroke only exists in an intermediate regime for the strength 
$B_s$. Clearly, the optimal performance shifts with increasing $S_p$ to
larger values of $B_s$ since a larger field is needed to move the
filament through the fluid. In other words, for larger $B_s$ the
filament is stiffer and a higher frequency $\omega \propto S_p^4$ is needed
to achieve the optimum stroke. Higher frequency means larger 
frictional forces and therefore a larger pumping performance, as
Figs. \ref{fig:magdepend}a) and b) demonstrate. When the magnetic
forces on the filament exceed the bending forces, one expects
the dynamics of the filament to be determined by the ratio of the  
hydrodynamic friction to magnetic forces, which we introduced in
Eq. (\ref{eq:Mason}) as Mason number $M_a$. From
Fig. \ref{fig:magdepend}a) we extract curves $\xi(S_p)$ for different
constant $B_s$, and rescale each of these curves with the respective
maximum values $\xi_m$. In Fig. \ref{fig:magdepend}c), all the data points 
for $B_s \ge 2$ are plotted as a function of the Mason number $M_a$. Indeed, 
most of the data points fall on a master curve, as predicted. Deviations 
occur for data points with $B_s$ close to 2.

\section{Pumping performance of hydrodynamically interacting cilia} 
\label{sec.multi}

Nature often uses arrays of beating cilia rather then a single isolated
cilium for generating fluid transport or to propel microorganisms such as
a paramecium \cite{Brennen77}.
These cilia beat in a synchronized fashion with a defined phase difference 
giving rise to so-called metachronal waves. As reviewed in the
introduction, analytical and numerical investigations of beating cilia 
suggest that
hydrodynamic interactions between single cilia synchronize their beating 
by inducing such a phase lag and therefore waves to occur.
For example, in Ref. \cite{Kim06} model cilia are driven by imposing a 
torque at the base of each cilium. This torque is reversed once the angular
amplitude of the cilium close to the base exceeds a certain threshold
value. Due to this mechanism, the phase of the driving torque and therefore
of each cilium can change during the beating cycle and a phase lag between 
neighboring cilia can evolve.
This cannot happen for the magnetically actuated cilia studied in this paper 
since the oscillating magnetic field determines the phase of each cilium. 
Nevertheless, one might ask if in our system the zero-phase-lag 
state is stable. We have never observed an instability towards a 
state with non-zero phase lag. 
This seems to be reasonable: the actuating magnetic field 
ultimately determines the forces which drive the filament against the
hydrodynamic friction through the liquid. Since these forces are much larger
than the hydrodynamic interactions between the cilia, a noticable phase lag
between the cilia cannot occur.
On the other hand, the artificial cilia system 
can be used to investigate in a systematic way how the pumping performance 
depends on the phase lag between the cilia 
by actuating each cilium with a separate magnetic field. In the following, 
we present initial results of such a study.

\begin{figure}
\includegraphics[width=1.2\columnwidth]{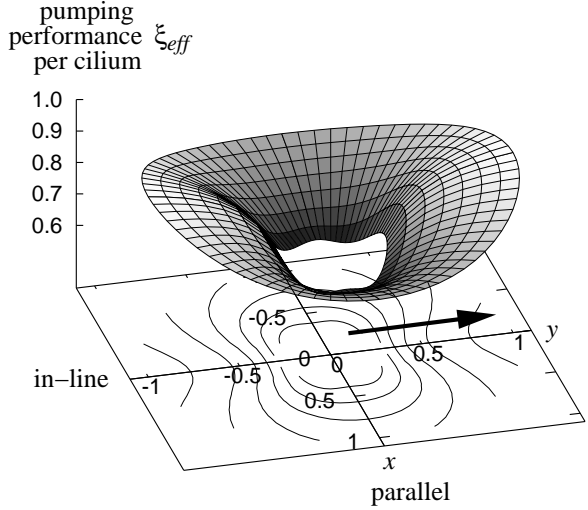}
\caption{Effective pumping performance $\xi_{eff}$ for a two-cilia system.
One cilium is placed at $x=y=0$, the other cilium at different
points in the $xy$ plane with coordinates in units of $L$. 
The effective pumping performance per cilium, $\xi_{eff}$, is given in 
units of the performance $\xi$ for a single, unperturbed filament 
actuated with the same parameters $S_{p} = 3$, $B_s = 3$  and 
$\varepsilon = 0.9$. The base contour lines run from $0.4$ (innermost) to 
$0.9$ (outermost) in steps of $0.1$. The arrow indicates the pumping
direction along the $y$ axis.}
\label{fig_multiperformance}
\end{figure}

In Fig.\ \ref{fig_multiperformance} we first investigate the pumping 
performance of a two-cilia system with zero phase lag between the
cilia. One cilium is placed at the center $x=y=0$ and the other at
different points in the $xy$ plane. We introduce an effective pumping
performance per cilium, $\xi_{eff}$, in units of the performance 
$\xi$ of a single cilium actuated with the same magnetic field cycle.
If the cilia are sufficiently well separated from each other, 
$\xi_{eff}$ approaches the pumping performance of a single filament, 
as expected. If the
cilia approach each other, $\xi_{eff}$ decreases. Due to hydrodynamic
interactions between the cilia beating with the same phase, their
effective friction with the surrounding fluid is reduced and therefore
they pump less fluid. Figure\ \ref{fig_multiperformance} clearly shows 
that this effect is more pronounced in the parallel configuration,
where the cilia are placed next to each other on the $x$ axis while
the strokes are performed in the $yz$ plane. On the other hand, in the
in-line configuration, where the cilia sit behind each other on the $y$ axis, 
$\xi_{eff}$ has almost reached the single-cilium performance $\xi$ at a 
distance $L$.

\begin{figure}
\includegraphics[width=0.8\columnwidth]{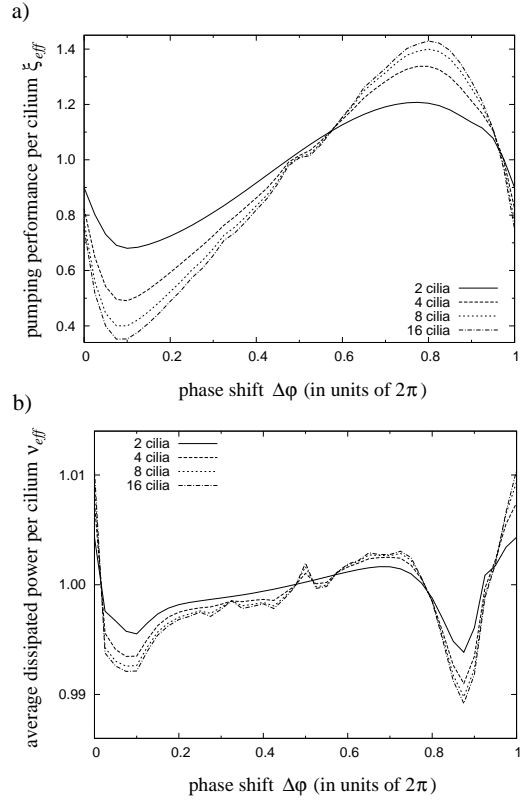}
\caption{a) Average effective pumping performance $\xi_{eff}$ for 
multi-cilia systems in the in-line configuration as a function of the 
phase shift between neighboring cilia. $\xi_{eff}$ is given in 
units of the performance $\xi$ for a single, unperturbed filament 
actuated with the same parameters $S_{p} = 3$, $B_s = 3$  and 
$\varepsilon = 0.9$.
b) Average dissipated power $\nu_{eff}$ for 
multi-cilia systems in the in-line configuration as a function of the 
phase shift between neighboring cilia. $\nu_{eff}$ is given in 
units of the dissipated power $\nu$ for a single, unperturbed filament.}
\label{fig_phaseplot}
\end{figure}

\begin{figure}
\includegraphics[width=0.8\columnwidth]{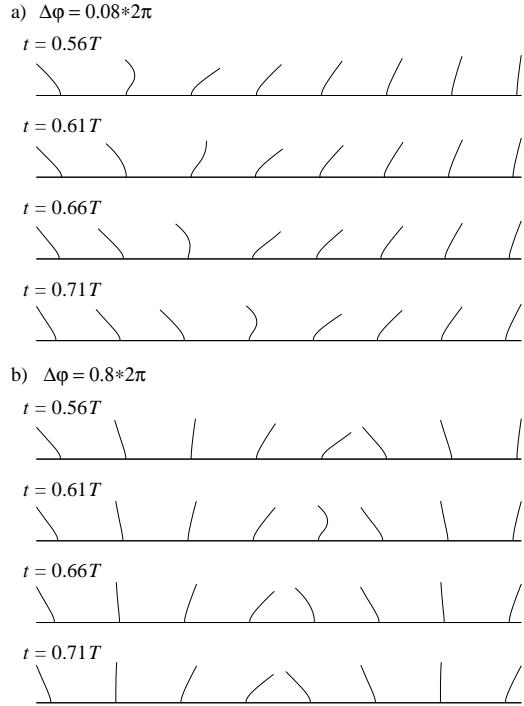}
\caption{Snapshots of the in-line configuration with eight cilia
for different phase shifts; a) $\Delta \varphi = 0.08 \times 2\pi$
(minimal pumping performance) and b) $\Delta \varphi = 0.8 \times 2\pi$
(optimum pumping performance). The times given in units of
the time period $T$ belong to the field cycle of the rightmost cilium.}
\label{fig_snapshotsII}
\end{figure}

Next we study the pumping performance and the dissipated energy of
several cilia placed on the $y$ axis in the in-line configuration.
Neighboring cilia are actuated by different magnetic fields with a 
prescribed phase shift $\Delta \varphi$. As illustrated by the first row in 
Fig. \ref{fig_snapshotsII}a), a cilium lags behind its neighboring cilium 
to the left by the small phase shift $\Delta \varphi$. Note that the slow 
transport stroke with a straight cilium goes to the right as can be seen 
from the snapshots at different times in Fig. \ref{fig_snapshotsII}a). 
Furthermore, when time proceeds, the strongly bent cilium in the recovery 
stroke also moves to the right. So metachronal wave propagation and 
transport stroke occur in the same direction which is also termed 
symplectic metachronism in literature \cite{Brennen77}. However, note 
that in biological systems the transport stroke is faster than the 
recovery stroke. On the other hand, for phase shifts close to 
$2 \pi$ [see Fig. \ref{fig_snapshotsII}b)] a cilium moves ahead of its 
neighbor to the left. As a result, the metachronal wave propagates 
opposite to the transport stroke and is therefore termed antiplectic. 
Figure \ref{fig_phaseplot}a) demonstrates 
that the average effective pumping performance per cilium, $\xi_{eff}$, 
falls below the reference value of a single cilium for small $\Delta \varphi$ 
while it assumes a maximum value at around $\Delta \varphi = 0.8\times 2\pi$, 
i.e., in the antiplectic mode. The maximal pumping performance $\xi_{eff}$
increases with the number of cilia and slightly moves to larger
$\Delta \varphi$. Even at the relatively large distance of $1.5L$ between 
the cilia, $\xi_{eff}$ is increased by more than $40 \%$  relative to
a single cilium. Preliminary calculations show that this value strongly
increases, when the cilia are moved closer together. Hence, our
study clearly demonstrates that metachronal coordination of ciliary 
beating at the right phase shift significantly enhances the ability to 
transport fluid. An explanation for this behavior can be inferred from 
the snapshots in Fig.\ \ref{fig_snapshotsII}. In the optimum stroke 
[see Fig.\ \ref{fig_snapshotsII}b)], the fifth cilium from left performs 
the recovery stroke against the neighboring fourth cilium which hinders 
the fluid flow initiated by recovery stroke and therefore increases 
fluid transport to the right. On the other hand, in the metachronal 
wave with lowest pumping performance [see Fig.\ \ref{fig_snapshotsII}a)], 
the cilium performing the recovery stroke is further away from the 
neighboring cilia and therefore its fluid transport is hardly hindered. 
While the pumping performance is strongly influenced by the phase shift 
and the number of beating cilia, the average dissipated power per 
cilium, $\nu_{eff}$, only exhibits very weak dependence on these parameters. 
As illustrated in Fig.\ \ref{fig_phaseplot}b), $\nu_{eff}$ deviates from 
the single-cilium value by at most 1 \%. This is in 
contrast to the work of Gueron and Levit-Gurevich, 
who, e.g., found that for a 10-cilia system with a distance of $0.7 L$
between the cilia $\nu_{eff}$ decreases by ca. 40 \% relative to the value 
of a single cilium \cite{Gueron99}. We checked that the smaller distance
is not the major reason for this difference. It might be due to the fact
that these authors model biological cilia with their internal 
actuation mechanism, whereas our cilia are actuated by an external field.

\section{Conclusion}

We have performed a thorough theoretical investigation of a magnetically 
powered low Reynolds number pumping device. It consists of a superparamagnetic 
elastic filament that is attached to a surface and that can be conveniently
actuated by an external magnetic field. We have introduced a reduced 
pumping performance $\xi$ that quantifies the fluid transport and compares 
it to the value of an idealized transport stroke of a rigid rod. With the help
of $\xi$, we have identified an optimum stroke pattern of the magnetically
actuated artificial cilium. It consists of a slow transport stroke and
a fast recovery stroke in contrast to nature. Our detailed 
parameter study for the reduced magnetic field $B_s$ and the sperm 
number $S_p$ reveals that for sufficiently large $B_s$ the behavior of the
artificial cilium is mainly governed by the Mason number that compares
frictional to magnetic forces. Initial studies on multi-cilia systems
reveal that the pumping performance is very sensitive to the imposed phase 
lag between neighboring cilia, i.e., to the details of the initiated
metachronal waves. Due to our results, we expect that metachronal waves 
in real cilia systems also increase the pumping performance for fluid 
transport.

Experiments using the superparamagnetic elastic filament for fluid 
transport are currently under way \cite{Fermigier08}. Our study will
help to identify relevant parameter ranges in which the artifical
cilium can be operated optimally. Initiating metachronal waves in the
multi-cilia system will be important for increasing the pumping
performance. This requires oscillating magnetic fields the phases
of which should vary from one cilium to the other. The realization of 
such a system is not unrealistic \cite{Maret08} but certainly poses a 
challenge to experimentalists. It would help to establish the 
artificial cilium as a useful tool for fluid transport in microfluidic
devices.

\begin{acknowledgement}
The authors would like to thank J. Bibette, R. Dreyfus, M. Fermigier, 
G. Maret, M. Reichert, and H. Stone for engaging and interesting discussions.
EMG and HS thank the International Graduate College at the 
University of Konstanz for financial support and the MPI for Dynamics
and Self-Organization in G\"ottingen for providing us with computing time.
EMG acknowledges support from the Marie Curie Early Stage Training 
network QIPEST (MEST-CT-2005-020505). MD and HS acknowledge financial 
support from the Deutsche Forschungsgemeinschaft under Grant No. Sta 352/7-1.
\end{acknowledgement}

\appendix

\section{Simulation Parameters} \label{app.par}

In table\ \ref{tab1}, we summarize the parameters used for the numerical 
simulations. The oscillation frequency $\omega = 2\pi/T$ and the 
magnetic-field strength $B$ were varied to study the respective ranges of 
sperm number $S_{p}$ and reduced magnetic field $B_s$. The spring 
constant $k$, the time step $\Delta t$ for the Euler integration, and
the number $n_{s}$ of simulation cycles were adjusted as necessary. Typical
values are shown.

\section{Mobilities} \label{app.wallmobilities}

Explicit expressions for the stresslet and sourcelet contricutions of 
Blake's tensor can be calculated from Ref. \cite{Blake71} or taken directly 
from Ref. \cite{Jones99}. Starting with these, we have calculated the 
self- and cross-mobilities as outlined in the main text.

\subsection{Self mobilities}

Due to the axial symmetry of the sphere with its image system all but the 
diagonal elements vanish and the $xx$ and the $yy$ elements are identical: 
\begin{equation}
\delta \bm{\mu}_{self} =  \frac{1}{6 \pi \eta a} \left( \begin{array}{ccc}
		\nu & 0 & 0 \\
		0 & \nu & 0 \\
		0 & 0 & 2 \nu
	\end{array}\right) ~,
\end{equation}
where $\nu$ is given by (with $a$ being the radius of the bead and 
$z$ the distance to the plane)
\begin{equation}
\nu = - \frac{3}{16} \left[\frac{a}{z} - \left( \frac{a}{z} \right)^3 
+ \frac{1}{3} \left( \frac{a}{z} \right)^5 \right] ~.
\end{equation}
Once the expression above is included into formula (\ref{selfm}) for
$\bm{\mu}_{ii}$, the self-mobility agrees with the expression given 
in Ref. \cite{Dufresne00}.

\subsection{Cross mobilities}

Symmetry dictates that the $xy$ and $yx$ elements are identical 
(the boundary wall is in the $xy$ plane). Furthermore, the $xx$ and $yy$ 
elements are formally equivalent when $R_x$ is interchanged with $R_y$. 
The same holds for the $xz$ and $yz$ or $zx$ and $zy$ elements, respectively.

The positions of the beads are $\bm{r}_1 = (x_1, y_1, z_1)$ and 
$\bm{r}_2 = (x_2, y_2, z_2)$ and the image system is located at  
$\bm{\overline{r}}_2 = (x_2, y_2, -z_2)$. Furthermore, for convenience 
we define $s = |\bm{r}_1 - \bm{\overline{r}}_2|$ and 
$R_x = (x_1 - x_2)$, $R_y = (y_1 - y_2)$ and $R_z = (z_1 + z_2)$. Let 
$\alpha$ and $\beta$ refer to the $x$ and $y$ but not the $z$ coordinate. 
In this notation, we obtain the following components for the 
cross mobilities:


\begin{figure*}
\begin{eqnarray}
(\delta \bm{\mu}_{12})_{\alpha \alpha}  & = &  \frac{1}{4 \pi \eta }  
\left[ -z_1z_2 \left( \frac{1}{s^3}-3{\frac { R_{\alpha}^2}{{s}^{5}}} \right)  
- \frac{a^2}{s^7} R_z^2 \left( 4 R_{\alpha}^2 - R_{\beta}^2 - R_z^2 \right) 
\right. \nonumber \\ 
& - & \left. \frac{a^4}{3 s^9} \left( 4R_{\alpha}^4 -R_{\beta}^4 + 
4 R_z^4 + 3 R_{\alpha}^2 R_{\beta}^2 + 3 R_{\beta}^2 R_z^2 - 
27 R_{\alpha}^2 R_z^2 \right) \right] \\
(\delta \bm{\mu}_{12})_{zz} & = & \frac{1}{4 \pi \eta } \left[ z_1z_2 
\left( \frac{1}{s^3}-3{\frac { R_z^2}{{s}^{5}}} \right)  - 
\frac{a^2}{s^7} R_z^2 \left( 3(R_{\alpha}^2 + R_{\beta}^2) - 
2 R_z^2 \right) \right.  \nonumber \\ 
& - & \left. \frac{a^4}{3 s^9} \left( 3  R_{\alpha}^4 + 3 R_{\beta}^4 + 
6 R_{\alpha}^2 R_{\beta}^2 - 24 R_z^2 (R_{\alpha}^2 + R_{\beta}^2) + 
8 R_z^4 \right) \right] \\
(\delta \bm{\mu}_{12})_{\alpha \beta} & =  & \frac{1}{4 \pi \eta } 
\left[ {\frac {3 z_1z_2 R_{\alpha} R_{\beta} }{{s}^{5}}} - 
\frac{5 a^2}{s^7} R_{\alpha} R_{\beta} R_z^2 - \frac{5 a^4}{3 s^9} 
\left( R_{\alpha}^2+ R_{\beta}^2 - 6R_z^2 \right) R_{\alpha} R_{\beta} 
\right] \\
(\delta \bm{\mu}_{12})_{\alpha z} & = &  \frac{1}{4 \pi \eta } 
\left[ R_{\alpha}  \left( {\frac {{z_1}^{2}}{{s}^{3}}}-
3{\frac {z_1z_2 R_z }{{s}^{5}}} \right)  +  R_{\alpha} a^2 
\left(\frac{1}{3} \frac{1}{s^3} + \frac{5}{s^7}R_z^3 - \frac{1}{s^5} 
\left[ R_z + 2(z_1^2 + z_1 z_2) \right] \right) \right. \nonumber \\
& + & \left. \frac{5 a^4}{3 s^9} R_{\alpha} (3 R_z [R_{\alpha}^2 +
R_{\beta}^2] - 4 R_z^3) \right] \\
(\delta \bm{\mu}_{12})_{z \alpha} & = &  \frac{1}{4 \pi \eta } 
\left[ R_{\alpha}  \left( {\frac {{z_1}^{2}}{{s}^{3}}}+
3{\frac {z_1z_2 R_z }{{s}^{5}}} \right)  +  R_{\alpha} a^2 
\left(\frac{1}{3} \frac{1}{s^3} - \frac{5}{s^7}R_z^3 + \frac{1}{s^5} 
\left[ R_z - 2(z_1^2 + z_1 z_2) \right] \right) \right. \nonumber\\
& - & \left. \frac{5 a^4}{3 s^9} R_{\alpha} (3 R_z [R_{\alpha}^2 
+R_{\beta}^2] - 4 R_z^3) \right]
\end{eqnarray}
\end{figure*}

\begin{table}
\caption{Parameters used in the numerical studies.}
\label{tab1}
\begin{tabular}{l@{\hspace{2cm}}l}
\hline
Parameter & Simulation value \\
\hline
\hline
$N$ & 20\\
$a ~ [\mu\mathrm{m}]$ &  0.5 \\
$l_{0}$  & 3$a$ \\[1ex] 
$\enspace\rightarrow L ~ [\mu\mathrm{m}]$ &  28.5\\
$\chi$ & 0.993 \\
$\eta ~ [\mathrm{Ns/m^2}]$ & $10^{-3}$ \\[1ex]
$k ~ [\mathrm{N/m}]$ & $1.5\cdot 10^{-3}$ \\
$A ~ [\mathrm{Nm}]$  & $4.5 \cdot 10^{-22}$ \\[1ex]
$\omega ~ [2\pi/s]\enspace$ & 0.071 \dots 140.72 \\
$\enspace\rightarrow S_{p}$ & 1.5 \dots 10 \\
$B ~ [\mathrm{T}]\enspace$ & 0 \dots 0.085 \\
$\enspace\rightarrow B_{s}$ & 0 \dots 7 \\
$\varphi_{\mathrm{max}} ~ [\enspace^{\circ}]$ & 60 \\[1ex]
$\Delta t ~ [\mathrm{s}]$ & $\mathcal{O}(10^{-6})$ \\
$n_{s}$ & $\approx$ 5 \\
\hline
\end{tabular}
\end{table}


\begin{thebibliography}{99}
\bibitem{Rotne}
In our modeling, we use a center-to-center distance $3a$ of the particles 
in the filament (where $a$ is the particle radius). In this case, the 
Rotne-Prager approximation deviates from the full multipole results of 
Ref.\ \cite{Cichocki94} by at most 5\%. We checked this by using the program 
Hydrolib (see Ref.\ \cite{Hinsen95}).
\bibitem{BVariation}
Another obvious way of achieving a non-reciprocal motion of the 
superparamagnetic filament would be to vary the magnetic field 
strength non-symmetrically during the actuation cycle. However, 
in our simulations we find that this method is less robust and effective.
\bibitem{footnote}
Note that formally the induced flow along the $y$ direction
should also depend on the force components $F_{xi}$ and $F_{zi}$.
Besides the fact that $F_{xi}=0$ in our case, by symmetry the integrals
of $G^{Blake}_{yx}$ and $G^{Blake}_{yz}$ over the $xy$ plane vanish
and additional contributions to Eq. (\ref{eq:delF}) do not occur.
\end{thebibliography}
\end{document}